\documentclass{ws-procs9x6}

\usepackage{subfigure}
\begin{document}

\title{Surrogates with random Fourier Phases}

\author{C. R\"ath, R. Monetti}

\address{Max-Planck Institut f\"ur extraterrestrische Physik\\
Giessenbachstr. 1,  85748 Garching, Germany}

\begin{abstract}

The method of surrogates is widely used in  the field of nonlinear data analysis 
for testing for weak nonlinearities. The two most commonly used algorithms for 
generating surrogates are the amplitude adjusted Fourier transform (AAFT) and 
the iterated amplitude adjusted Fourier transfom (IAAFT) algorithm. Both the AAFT 
and IAAFT algorithm conserve the amplitude distribution in real space and 
reproduce the power spectrum (PS) of the original data set very accurately. \\
The basic assumption in both algorithms is that higher-order correlations can 
be wiped out using a Fourier phase randomization procedure. In both cases, 
however, the randomness of the Fourier phases is only imposed before the 
(first) Fourier back tranformation. Until now, it has not been studied how the 
subsequent  remapping and iteration steps may affect the randomness of the 
phases.\\
Using the Lorenz system as an example, we show that both algorithms 
may create surrogate realizations containing Fourier phase correlations. We present 
two new iterative surrogate data generating methods being able to control the 
randomization of Fourier phases at every iteration step. The resulting surrogate 
realizations which are truly linear by construction display all properties needed for 
surrogate data. 

\end{abstract}


\bodymatter
\section{Introduction}
The detection of non-linear behavior in data sets is a general problem which arises in diverse 
disciplines. \cite{CC,CN,CNC,MAT,MAT1,Raeth03,BA,LE,SI}. 
Tests for non-linearity which compare a data set to the null hypothesis of a 
Gaussian linear process are relevant either to constrain models or support theories 
of a particular system. In this context, the surrogate data tests are model independent 
tests for non-linearity \cite{Schreiber96,SA,SR,Theiler92} 
which can be applied with any 
non-linear statistics that characterizes a data set. 
Ideal surrogate data should possess not only the same power spectrum and 
amplitude distribution in real space as the original data set but also be free 
of higher-order correlations. The Amplitude Adjusted Fourier Transform (AAFT) 
and the Iterative Amplitude Adjusted Fourier Transform (IAAFT) 
algorithms \cite{Schreiber96,SR,Theiler92} 
are the most popular algorithms to generate an ensemble of surrogate realizations.
AAFT and IAAFT algorithms conserve the amplitude distribution in real space and 
reproduce the power spectrum (PS) of the original data set quite accurately. 
The basic assumption in both algorithms is that higher-order correlations can be 
wiped out using a Fourier phase randomization procedure since the PS is invariant 
under Fourier phase permutations. Previous studies \cite{Raeth02,Schreiber96,DK} 
focused on the ability of surrogate generating algorithms to accurately reproduce 
the PS of the original data. However, it has been shown that for an ARMA process 
an ensemble of surrogates generated by the AAFT algorithm displays an anomalous 
small variability of the PS when compared to an ensemble of identical 
ARMA processes. \cite{Dolan01}. 
This poses the question of the definition of the null hypothesis for the surrogate 
test which will certainly depend upon the system under study. 
More important than exactly reproducing the PS of the original data 
is to create surrogate data free of higher-order correlations 
which by definition are linked to the properties of the Fourier phases. 
However, how phase correlations are related to higher-order 
correlations is an important yet unresolved open problem. 
AAFT and IAAFT algorithms impose constraints on the Fourier 
amplitudes and the amplitude distribution in real space without controlling 
Fourier phases. As shown in the next section, 
these constraints may generate Fourier phase correlations.
In the rest of this contribution we propose and discuss novel 
surrogate generating algorithms in which the randomness of the 
phases is explicitly controlled.

\section{Phase correlations in surrogates}

In this section we introduce the model systems which we use in our study.
Furthermore we demonstrate, how phase correlations, thus non-linearities,
can be induced in surrogate data sets.
\subsection{Model systems}
As a linear system we consider the first order ARMA process described by
\begin{equation}
x_{n+1}=0.7x_n + \xi_n \;,
\end{equation}
with $\xi_n$ being Gaussian noise.\\
Furthermore, we use the $z$-component of the  Lorenz system \cite{LO}
\begin{eqnarray}
\dot{x}& =&  a(y-x)\\
\dot{y}& =&  x(b-z) - y\\
\dot{z}& =&  xy-cz
\end{eqnarray}
in the chaotic regime , i.e. $a=10$, $b=28$ and $c=8/3$,
as an example for a time series derived from a 
deterministic chaotic system, which explicitly 
contains non-linear correlations.
The original data were sampled at $\tau'=0.001$ and a rank ordered 
remapping onto a Gaussian distribution was performed before applying the surrogate algorithms. 
The original time series resulting from the rank ordered remapping onto a Gaussian distribution 
is called $z_L$ throughout the text.  The length of the times series is $M=2^{15}$.

\begin{figure}[h]
\begin{center}
\includegraphics[width=5.5cm,angle=0]{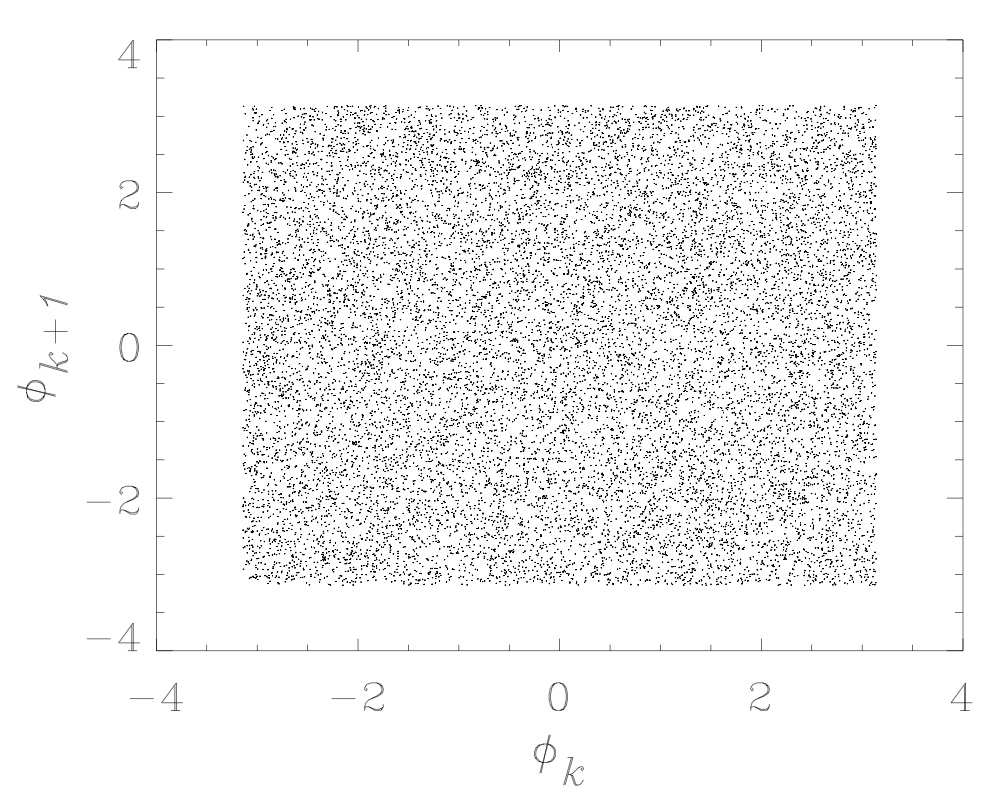}
\includegraphics[width=5.5cm,angle=0]{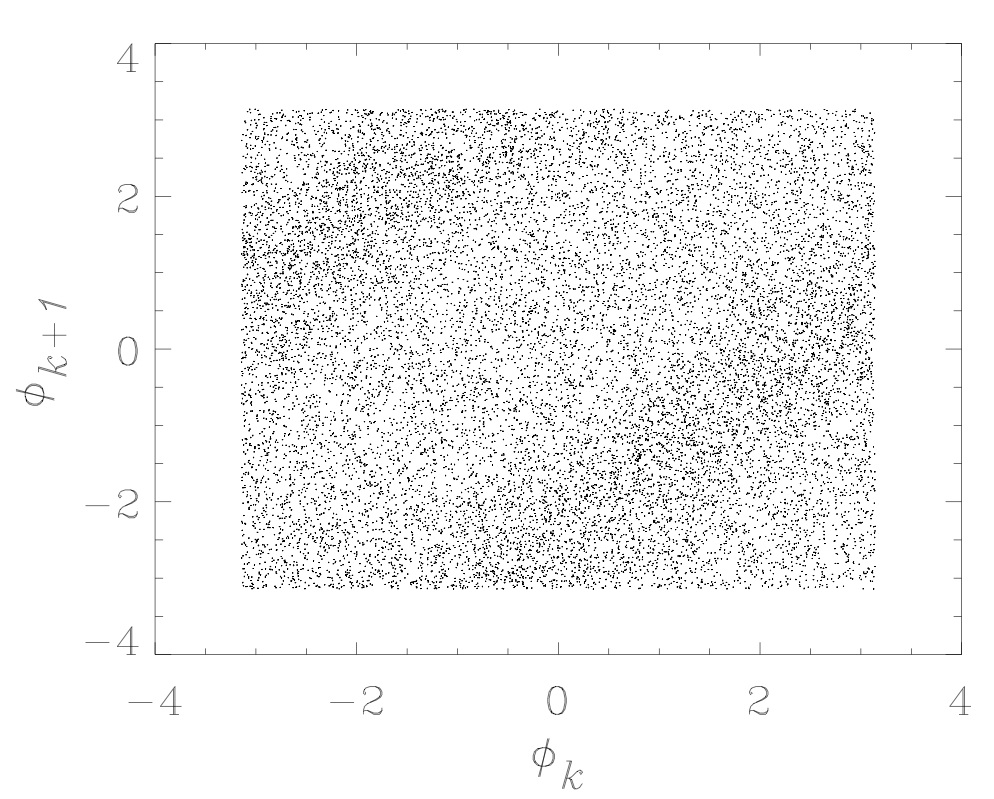}
\includegraphics[width=5.5cm,angle=0]{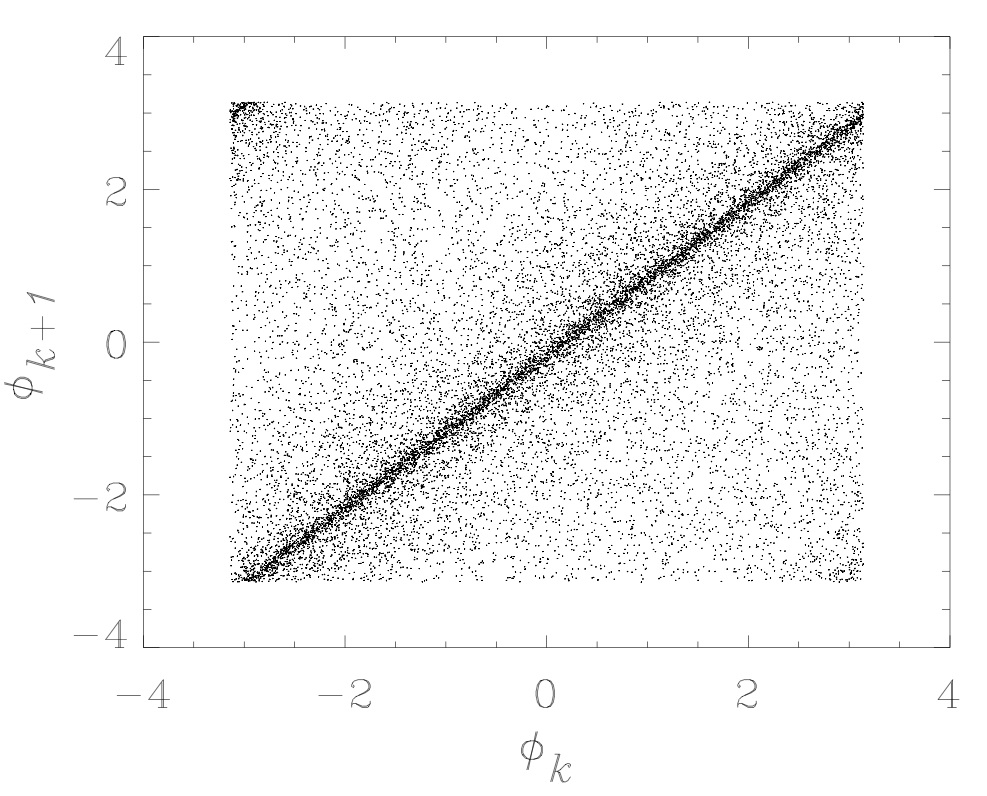}
\includegraphics[width=5.5cm,angle=0]{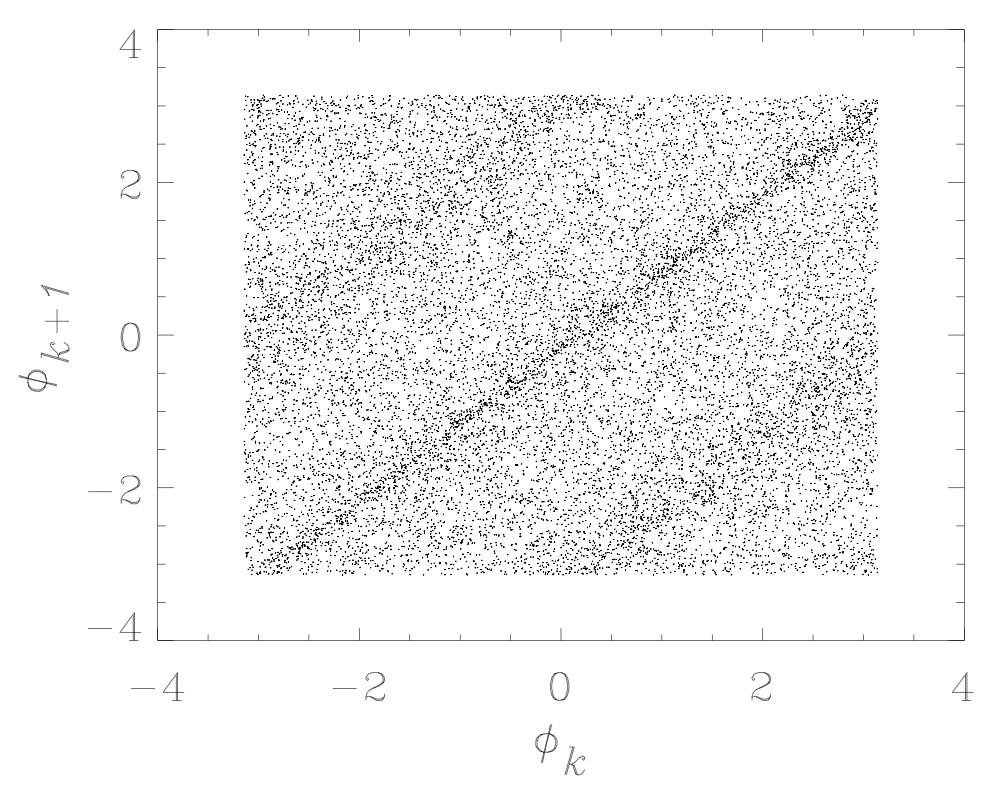}

\caption{Phase maps for the linear ARMA process and its surrogates (for details see text). }
\label{figure1}
\end{center}
\end{figure}

\begin{figure}[h]
\begin{center}
\includegraphics[width=12cm,angle=0]{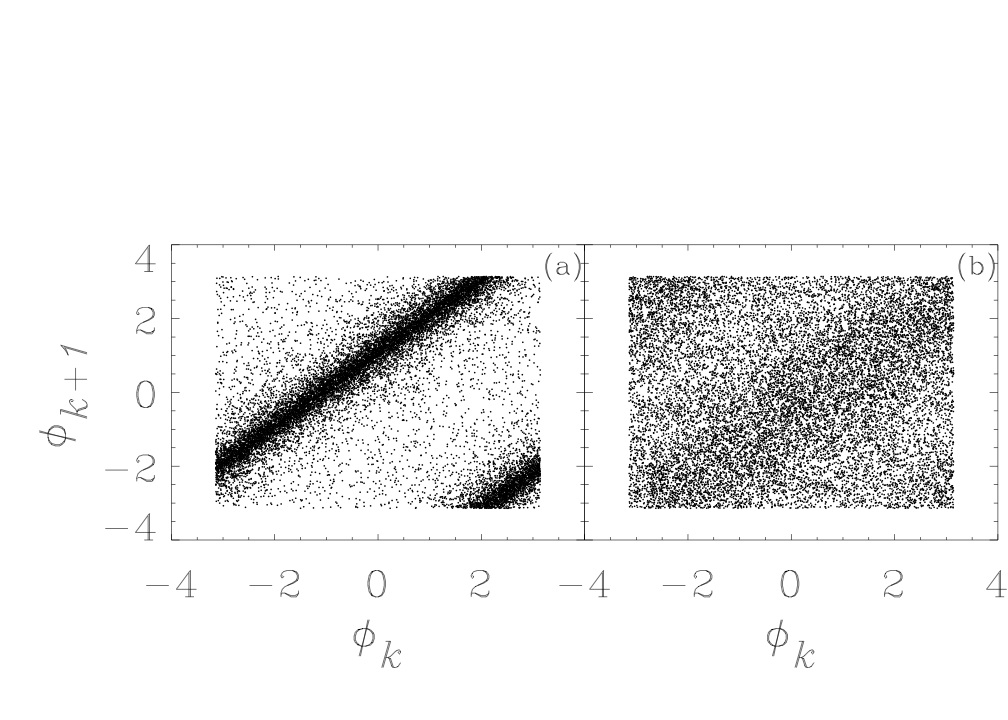}
\caption{ a) Fourier phase map of an IAAFT surrogate realization of  $z_L$ for $\Delta=1$. 
b) same as (a) but using an AAFT surrogate realization.}
\label{figure2}
\end{center}
\end{figure}

\subsection{Phase maps} 
A Fourier phase map is a two-dimensional set of points $G=\{ \phi_k,\phi_{k+\Delta} \}$ where $\phi_k$ is the phase of the $k^{th}$ mode of the Fourier transform and $\Delta$ a phase delay. All possible modes $k$ up the Nyquist frequency are considered. Using this representation, phase correlations can show up in the maps as high/low density regions. 

\subsection{Inducing phase correlations}

First, we consider the ARMA process described above. 
The amplitude distribution in real space for the ARMA process is Gaussian. 
Figure \ref{figure1} upper left panel shows a Fourier phase map for a relization of this ARMA process. 
As expected, this phase map shows no signatures of phase correlations. 
Suppose that we replace two data point by two spikes thus the distribution is almost Gaussian. 
Figure  \ref{figure1} upper right panel shows the Fourier phase map for the spiky time series. 
It is clear that the presence of spikes has induced Fourier phase correlations. 
Now, we generate an IAAFT surrogate realization of the spiky times series. 
Figure  \ref{figure1} lower left panel shows the Fourier phase map for this IAAFT surrogate where once 
again the Fourier phases are strongly coupled. The first guess is that this strong coupling 
in the phases is only due to the spikes still present in the IAAFT realization and 
that this will vanish after removing the spikes. However, as shown in Fig.  \ref{figure1} 
lower right panel, phases correlations still remain after removing the spikes. 
Then, this residual phase correlations have been induced during the iterative process 
leading to an IAAFT surrogate which displays signatures of non-linearity. 
We conclude that Gaussian remapping must always be performed since this step 
warrants that we obtain random uniformly distributed Fourier phases 
at the first iteration step of IAAFT.\\ 
Second, we calulate AAFT and IAAFT surrogate realisations of the (Gaussian remapped) 
$z$-component of the Lorenz system $z_L$ and represent the phase information using phase 
maps (Fig.  \ref{figure2}). It becomes immediately obvious that both surrogate generating 
algorithms can induce phase correlations.

\section{The IPAFT algorithm}
We introduce two alternative iterative methods to generate surrogate realizations 
where the randomization of higher-order correlations is controlled at every iteration 
step by imposing uncorrelated uniformly distributed random Fourier phases. 
For data having a Gaussian amplitude distribution in real space, the resulting 
surrogate realization will have a similar distribution in real space, a well-reproduced 
auto-correlation function, and uncorrelated uniformly distributed Fourier phases.

Algorithm A consists of the following steps. Consider a time series $y_n, \, n=1, \ldots ,M$. (i) Perform a rank-ordered remapping of the original data onto a Gaussian distribution $x_n = G(y_n)$. 
Remapping a time series onto a Gaussian distribution conserves dynamic non-linearities \cite{Theiler92} and reduces the amount of whitening of the PS induced by the iterations. 
Evaluate the Fourier transform of $x_n$. In addition, generate a random shuffling realization of  $x_n$ and calculate its Fourier transform. Since the amplitude distribution in real space is Gaussian, we obtain uncorrelated uniformly distributed Fourier phases in the interval $[-\pi, \pi]$.  If we avoid the remapping step onto a Gaussian distribution then the shuffling step cannot ensure uncorrelated uniformly distributed Fourier phases as shown above \cite{MAT,MAT1}. (ii) Combine the Fourier amplitudes of  $x_n$ with the random Fourier phases and perform an inverse Fourier  transformation. Let us call the resulting time series $x'_n$. (iii) Sort $x_n$ as $x'_n$ ($z_n = S_{x'_n}(x_n)$) and evaluate the Fourier transform of $z_n$. Steps (i), (ii), and (iii) describe the AAFT algorithm, i.e. $z_n$ is an AAFT surrogate realization. 
Instead of replacing the Fourier amplitudes of $z_n$ by the original Fourier amplitudes, as is the case for IAAFT, we replace the Fourier phases by the random phases obtained in step (i). Using the new Fourier amplitudes and the random phases we repeat steps (ii) and (iii). Iterations stop after convergence of the probability distribution in real space. In the last iteration, step (iii) is avoided thus we obtain a surrogate realization containing uncorrelated uniformly distributed Fourier phases.
Step (iii) may actually reintroduce uncontrolled Fourier phase correlations as shown in Fig. \ref{figure2} (b). 
Imposing random Fourier phases at every iteration step will prevent the algorithm to develop phase correlations during evolution and so it maintains the randomization of higher-order correlations. However, the original PS is imposed only at the first iteration[s] and evolves freely through subsequent iterations. This freedom given to Fourier amplitudes introduces, however, only small deviations to the PS after the first iteration step, which is the one that brings the initial flat PS to the desired one. Convergence was verified in several tests even in higher dimensions. It is expected since a Gaussian linear process is characterized by both a Gaussian amplitude distribution and uncorrelated uniformly distributed Fourier phases.  

Using the ARMA process, we quantified the spectral variability of an ensemble of 
phase-controlled surrogates via $\alpha=\frac{1}{N} \sum_{k=1}^{N} \frac{(P_k - <P_k>)^2}{<P_k>^2}$, 
where $P_k$ is the power of the $k$-mode of a surrogate realization and $<P_k>$ is 
the mean power \cite{Dolan01}. The spectral variability is 
compared with the actual variability of an ensemble of ARMA processes using  $<\alpha>_S$ and $\sigma_S(\alpha)$. Our results (see Table below) show that in spite of allowing for freely evolving Fourier amplitudes, the spectral variability of phase-controlled surrogates is almost two orders of magnitude smaller than that of the ensemble of ARMA processes.
Algorithm B tackles this problem by allowing a variance of $P_k$ over the set of surrogates given by $\sigma_k^2 = P^2_k$ \cite{OS}. It is identical to Algorithm A except for introducing the variance in step (ii) only at the first iteration. 

\begin{table}[h]
\begin{center}
\begin{tabular}{|c|| c| c|} 
\hline
Ensemble & $< \alpha >_S$ & $\sigma_S(\alpha)$  \\
\hline\hline
ARMA process & 1.04 & 0.03   \\ 
Phase-controlled A & 0.016 & 0.006  \\ 
Phase-controlled B & 0.67 & 0.02 \\ 
\hline
\end{tabular}
\caption{Spectral variability. Results were obtained using 100 realizations. 
                A and B stand for algorithms A and B, respectively.}
\label{tab1}
\end{center}
\end{table}

\section{Results}

\subsection{Convergence}

We consider the convergence of the amplitude distribution in real space 
as the stopping criterion for the algorithms. Convergence is assessed 
via the relative deviation of the amplitude distribution defined 
as $\Delta I^{(i)}=\sum_{k=0}^{M-1} (r_k^{(i)} -r_k)^2 / \sum_{k=0}^{M-1} r_k^2$, 
where $r_k^{(i)}$ is the rank-ordered time series resulting after step (ii) at 
iteration $i$, and $r_k$ is the original rank-ordered time series. By skipping the 
last rank-ordered remapping step in AAFT and IAAFT algorithms, 
it is possible to compare the performance of all algorithms to reproduce the amplitude distribution in real space. 
The Table above shows the results for $\Delta I^{(\infty)}$, i.e. the value of $\Delta I^{(i)}$ after convergence. This table
indicates that phase-controlled surrogates reproduce the amplitude distribution in 
real space one order (six orders) of magnitude better than IAAFT (AAFT), respectively. 

\begin{table}[!t]
\caption{Mean value and standard deviation of the relative deviation $\Delta I^{(\infty)}$ for all surrogate classes. Results were obtained using 50 surrogate realizations.}
\label{tab2}
\begin{center}
\begin{tabular}{@{} |l|| @{\hspace{5mm}} c| c|} 
\hline
\rule[-1ex]{0pt}{3.5ex}  Ensemble & $< \Delta I^{(\infty)} >_S$ & $\sigma_S(\Delta I^{(\infty)})$  \\
\hline\hline
\rule[-1ex]{0pt}{3.5ex} AAFT & 0.0054 & 0.0028   \\
\rule[-1ex]{0pt}{3.5ex} IAAFT & 5.9x10$^{-7}$ & 4.4x10$^{-7}$ \\ 
\rule[-1ex]{0pt}{3.5ex} Phase-controlled A & 1.7x10$^{-8}$ & 1x10$^{-9}$  \\ 
\rule[-1ex]{0pt}{3.5ex} Phase-controlled B & 1.7x10$^{-8}$ &  1.1x10$^{-9}$ \\ 
\hline
\end{tabular}
\end{center}
\end{table}

\begin{figure}
\begin{center}
\includegraphics[width=12cm,angle=0]{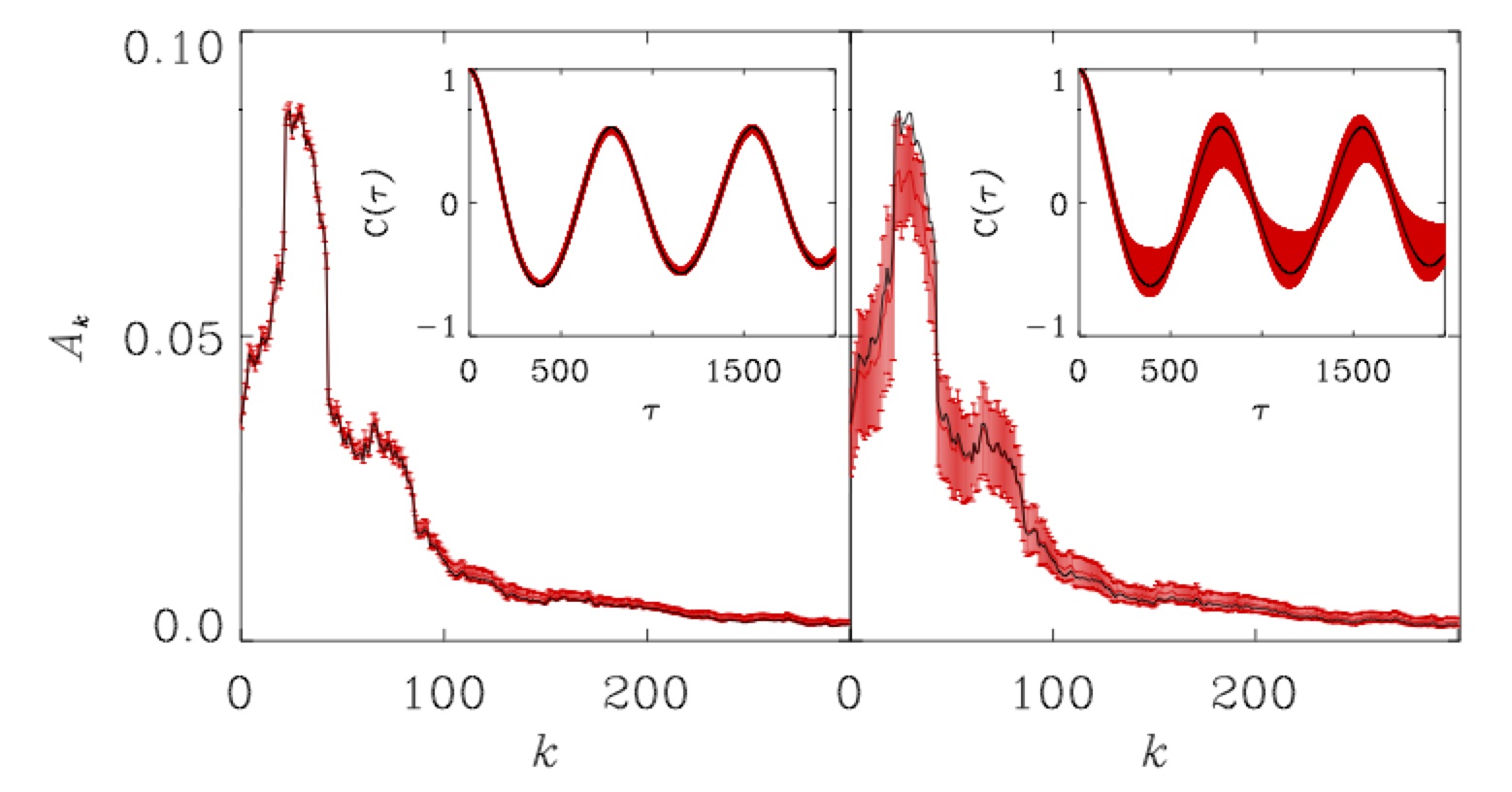}
\caption{(left) Smoothed Fourier amplitude $A_k$ versus the wave number $k$
for $z_L$. The inset shows the autocorrelation function $C(\tau)$ versus the 
time lag $\tau$ for $z_L$. The shaded region shows 1$\sigma$ error 
bars obtained using 50 phase-controlled surrogate realizations generated 
by algorithm A. (right) Same as  the left figure but using 50 phase-controlled surrogate 
realizations generated by algorithm B. }
\label{figure3}
\end{center}
\end{figure}

Figure \ref{figure3} shows $A_k$ versus $k$ and $C(\tau)$ versus $\tau$ for the $z$ 
component of the Lorenz system $z_L$
for algorithms A and B after convergence. These results indicate that 
both algorithms lead to well-behaved surrogate realizations. 
A comparison of Figs. \ref{figure3} (left) and \ref{figure3} (right) shows 
that introducing a variance in the PS leads to a widening of the 1$\sigma$ error bars. 

\begin{figure}
\begin{center}
\includegraphics[width=12cm,angle=0]{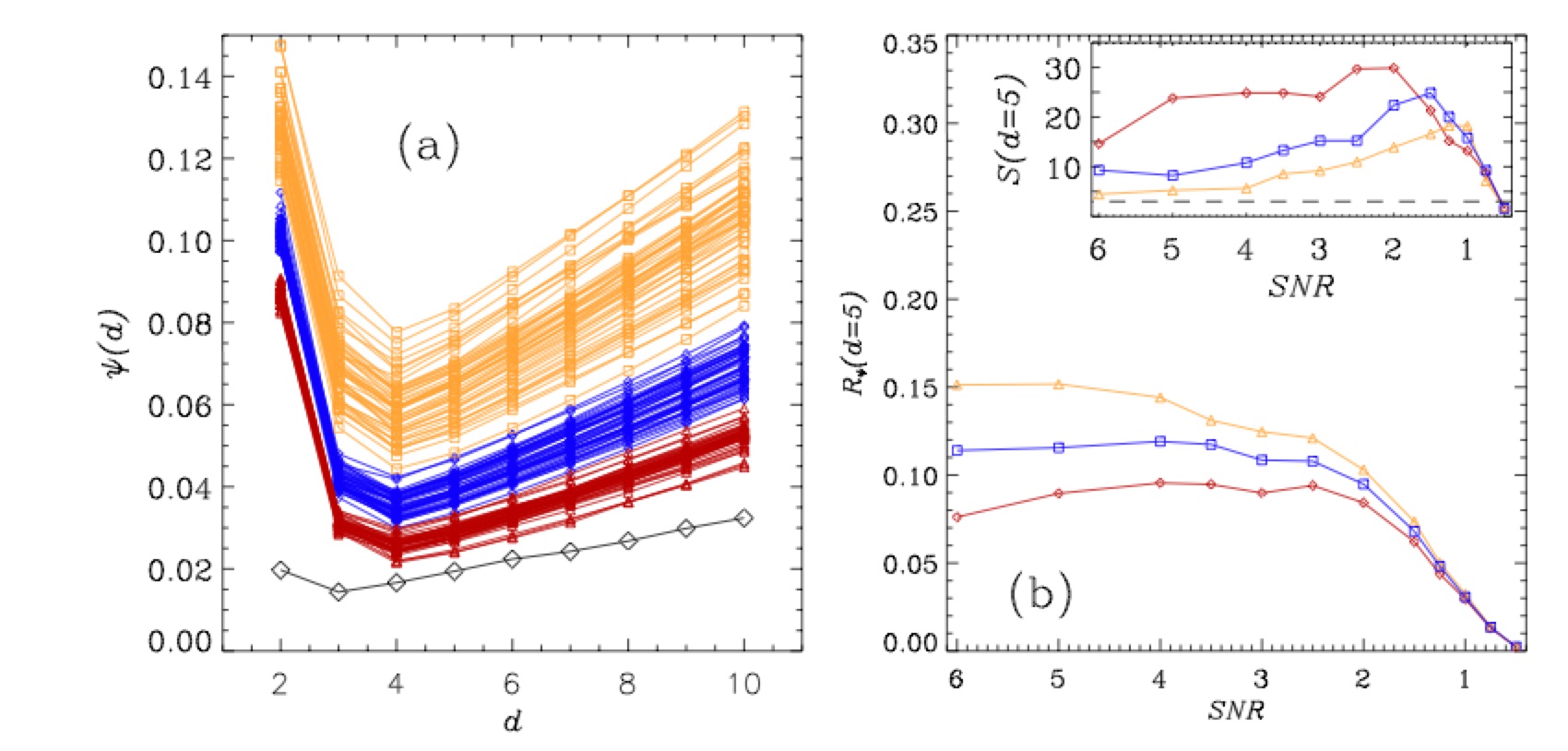}
\includegraphics[width=12cm,angle=0]{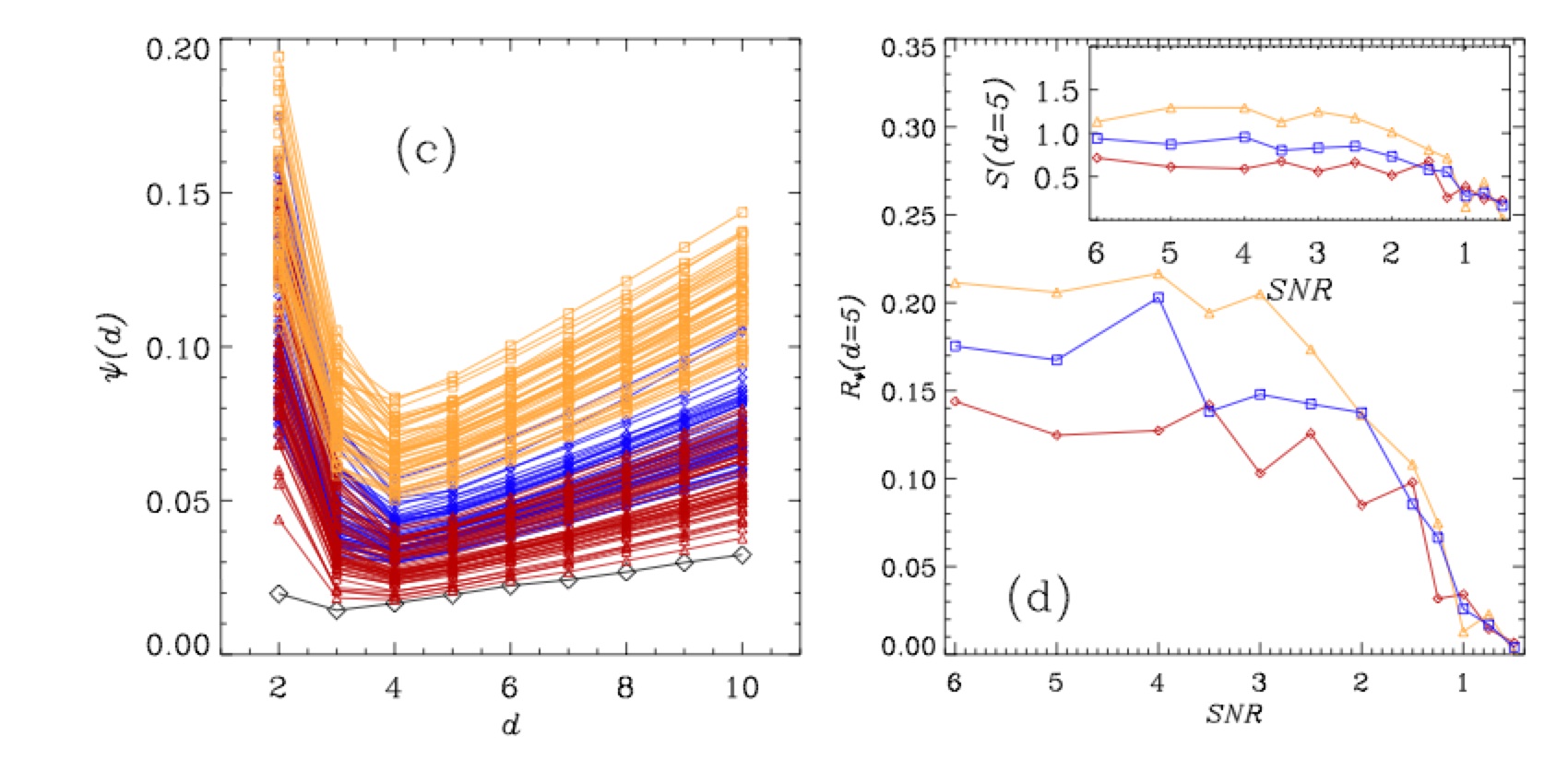}
\caption{a) $\psi$ versus the embedding dimension for $z_L$ (black curve), an ensemble of IAAFT (red) , 
AAFT (blue), and phase-controlled (A)  (yellow) surrogate realizations.
 b) The relative deviation $R_{\psi}(d=5)$ versus $SNR$ for the three classes. 
 $R_{\psi}(d=5) = |\psi_O -<\psi>_S|/<\psi>_S$ where $\psi_O$ is the NLPE value for the original data and $<\psi>_S$ is the mean value
of $\psi$ as derived from surrogate realizations, for embedding dimension $d=5$. The inset shows the $\sigma$-normalized deviation versus $SNR$ for the three classes. $S(d)=|\psi_O (d) -<\psi>_S|/\sigma_S(\psi)$, where $\sigma_S(\psi)$ is the standard deviation of $\psi$ as derived from surrogate realizations.
c) and d) The same as  a) and b) but using phase-controlled (B) surrogate realizations, respectively.}
\label{figure4}
\end{center}
\end{figure}

\subsection{Applications to the Lorenz-System}

We performed a comparative study of the four different surrogate generating algorithms, 
namely AAFT, IAAFT, phase-controlled A, and phase-controlled B surrogates. 
To this purpose, we applied the non-linear prediction error (NLPE) 
method \cite{SR,NLPE} to test for non-linearity.  
To calculate NLPE, the time series is embedded in a $d$-dimensional space 
using the method of delay
 coordinates: $\vec{x}_n= (x_{n-(d-1) \tau}, x_{n-(d-2)\tau}, \ldots , x_n )$, 
 where $\tau$ is the delay time. Then, we define the NLPE as
\begin{equation}
\psi(d,\tau,T,N) = \frac{1}{(M-T-(d-1)\tau)} \Big( \sum_{n=(d-1)\tau}^{M-1-T} [\vec{x}_{n+T} -F(\vec{x}_n)]^2 \Big)^{1/2} 
\end{equation}
where $F$ is a locally constant predictor, $M$ is the length of the time series, 
and $T$ is the lead time. The predictor $F$ is calculated by averaging over future values 
of the $N \, (N=d+1)$ nearest neighbors in the delay coordinate representation. 
We have studied the behavior of $\psi$ as a function of the lead time $T$. 
We found that for $T > 5$ $\psi$ remains rather constant, 
thus a value of $T=10$ was used for this test.
Using $\psi$, we evaluated the $\sigma$-normalized deviation $S$ and 
the relative deviation $R$ (see caption of  Fig. \ref{figure4}). 
In all cases, $\tau$ was chosen in order to satisfy the criterion 
of zero autocorrelation \cite{OEM}. 
Figure \ref{figure4}(a) shows $\psi$ versus the embedding dimension $d$ for IAAFT, AAFT and phase-controlled A surrogates. NLPE distinguishes between the three groups of surrogates and indicates that both AAFT and IAAFT surrogate realizations are more predictable than phase-controlled A surrogates, i.e. $\Big\| \psi_O -<\psi>_S \Big\| $ is larger for phase-controlled A surrogate realizations for all embedding dimensions. It is remarkable that the three groups do not overlap. Figure \ref{figure4}(a) also shows that the largest variability is observed for the set of phase-controlled A surrogate realizations and it decreases for AAFT and IAAFT surrogate realizations, respectively.\\ 
The $\sigma$-normalized deviation, which is a function of the surrogate ensemble variability and $\Big\| \psi_O -<\psi>_S \Big\| $, takes the values $S(5) = 5.1, \, 6.9, \, \text{and} \, 5.8$ for IAAFT, AAFT, and phase controlled A surrogates, respectively.

We also analyzed the behavior of all surrogate classes in the presence of noise. We took $z_L$ and superimposed additive Gaussian white noise with varying standard deviation $\sigma_n$. The signal to noise ratio is defined as $SNR=\sigma_s/\sigma_n$ where $\sigma_s$ is the standard deviation of the original time series. 
In the presence of noise, the behavior of $\psi$ versus $d$ resembles that of Fig. \ref{figure4}(a) but differences among the three surrogate classes are now smaller. The inset plot of Fig. \ref{figure4}(b) shows $S(5)$ versus $SNR$ for IAAFT, AAFT and phase-controlled A surrogates. For $SNR \le 0.5$, the three groups lead to non-significant results as expected. For $SNR > 0.5$, $S(5)$ grows rapidly and displays exceptional high values for all surrogate classes  for relatively low $SNR$ values. In fact, maxima values of $S(5) \sim 20, 25, \, \text{and} \, 30$ are observed at $SNR \sim 1, 1.5, \, \text{and} \, 2$ for phase-controlled A, AAFT, and IAAFT surrogates, respectively. All surrogate classes display a counterintuitive decreasing behavior when further increasing $SNR$.  Even for the highest $SNR$ value here studied ($SNR=6$), IAAFT and AAFT display $S(5)$ values higher than the value obtained in the absence of noise. However, for phase-controlled A surrogates this feature is only observed for $SNR < 3.5$. 
Two different effects are responsible for this trend, namely the increase of $|\psi_O -<\psi>_S|$ and the decrease of $\sigma_S(\psi)$ with decreasing $SNR$. 
However, this trend is more pronounced for IAAFT and AAFT surrogates than  due to the smaller variability observed for the standard surrogate data which already becomes clear in Fig. \ref{figure4} (a). $\sigma_S(\psi)$ is approximately one order of magnitude (5 times) smaller for IAAFT (AAFT) surrogates than for phase-controlled A surrogates. For $SNR \sim 0.5 $, $|\psi_O -<\psi>_S|$ sharply drops to zero thus yielding the expected non-significant results for all surrogate classes. 
Fig. \ref{figure4}(b) shows the relative deviation $R_{\psi}(5)$ versus $SNR$. Note that the order of the curves has been mirrored with respect to their order in the inset plot. Since $S=\frac{R_{\psi} <\psi>_S}{\sigma_S(\psi)}$, then $\sigma_S(\psi)$ is responsible for this reordering. 
$R_{\psi}(5)$ not only shows the expected trend for increasing $SNR$ but it also takes larger values when evaluated in the absence of noise ($R_{\psi}(5) = 0.33, 0.51, \, \text{and} \, 0.70$ for IAAFT, AAFT, and phase-controlled A surrogates respectively). Thus, it provides a suitable relative measure of significance for different $SNR$. 
This behavior is also observed for other embedding dimensions. 

Figure \ref{figure4}(c) shows $\psi$ versus $d$ for IAAFT, AAFT and phase-controlled B surrogates. In order to allow for a fair comparison among surrogate classes, we modified both the IAAFT and AAFT algorithms by introducing a variability in the PS as done in algorithm B.
Phase-controlled B surrogates are still distinguished as a single group by the NLPE and display the largest variability. In the absence of additive noise phase-controlled B surrogates lead to the most significant result $S(5) = 5.9$ followed by AAFT ($S(5) = 3.6$), while IAAFT is non-significant ($S(5) = 2.1$). The inset plot of Fig. \ref{figure4}(d) shows that in the presence of noise the $\sigma$-normalized deviation has dropped well below the 3$\sigma$ threshold for all surrogate classes. However, it displays the expected increasing trend with $SNR$ values thus providing an equivalent description as the relative deviation $R_{\psi}(5)$. 

\begin{figure}
\begin{center}
\includegraphics[width=12cm,angle=0]{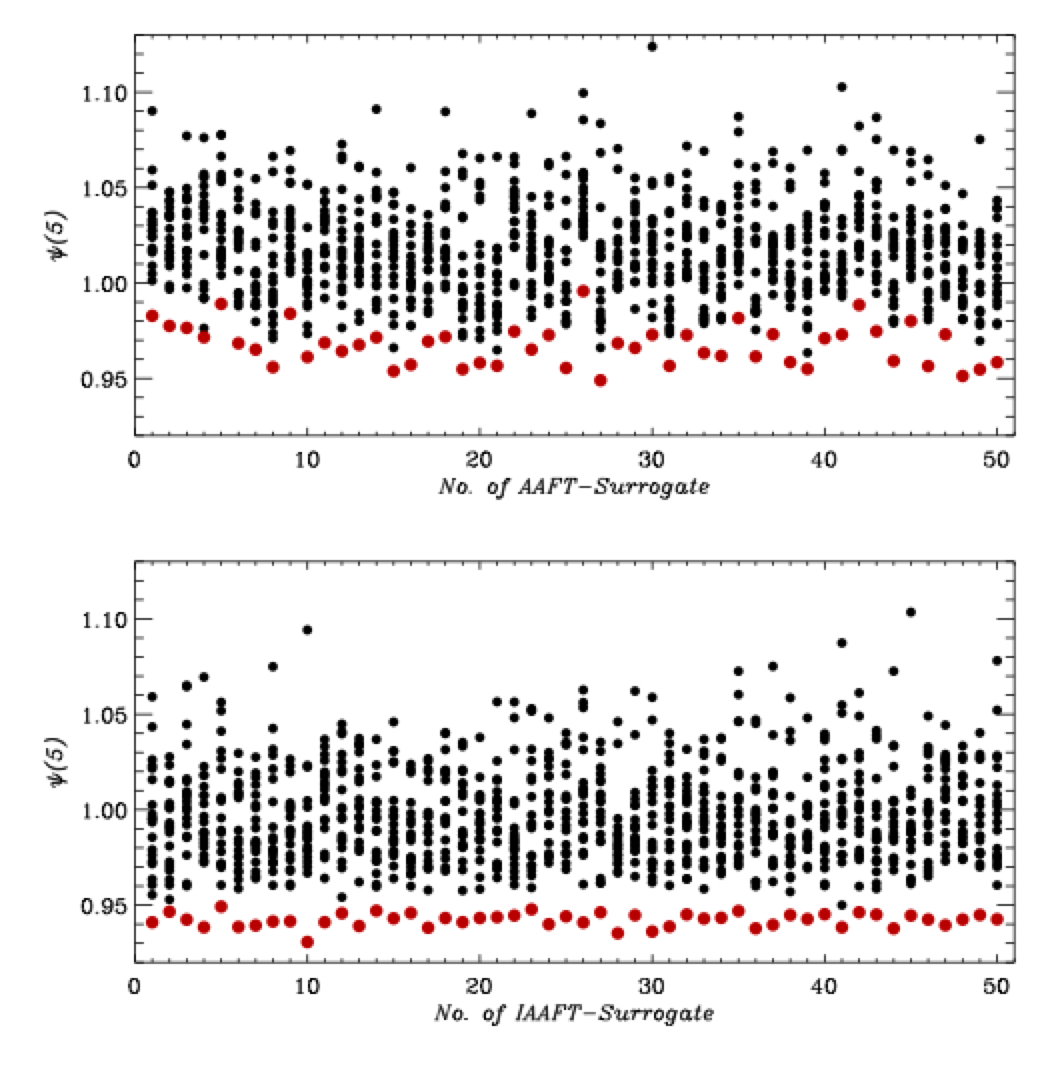}
\caption{ $\psi(5)$ values for 50 AAFT (above) and 50 IAAFT (below) surrogate realizations of $z_{L}$ 
for $SNR = 4$ (red dots) and for 20 phase-controlled (A) surrogates generated from each of them 
(black dots). 
 }
\label{figure5}
\end{center}
\end{figure}

To investigate the differences between surrogate classes, we considered all the 50 AAFT and 50 IAAFT surrogate realizations of $z_L$ for $SNR=4$ and generated 20 phase-controlled A surrogates for each of them. Red dots in Figure \ref{figure5} show the values of $\psi(5)$ for AAFT and IAAFT surrogate realizations and black dots show the values of $\psi(5)$ for the respective phase-controlled A surrogates.  We observe that for both AAFT and IAAFT surrogates, phase-controlled A surrogates lead to larger $\psi(5)$ values in all cases. It is worth noting that in all $2 \cdot 50 = 100$ tests, none of the phase-controlled A surrogate realizations showed lower (more predictable) values of $\psi(5)$ than the respective AAFT (IAAFT) surrogate.
These differences in $\psi(5)$ between standard surrogates and phase-controlled A surrogates generated from them become even larger for higher $SNR$ values. Since algorithm A accurately reproduces the PS of the original data and phase correlations in standard surrogates become stronger for higher values of $SNR$, we attribute the differences unveiled by the NLPE to the presence of Fourier phase correlations. 

\section{Conclusions}
We demostrated that the AAFT and IAAFT algorithm may generate 
surrogate realizations containing Fourier phase correlations. 
Motivated by this finding,  we presented two new surrogate generating algorithm 
being able to control the randomization of Fourier phases at every iteration step. 
Using the NLPE method we found clear  differences among the surrogate 
classes that we attribute to the randomization level of the Fourier phases 
of the surrogate realizations.  
These differences appear due to the constraints imposed on the surrogate generating procedure. 
As shown above, freely evolving Fourier amplitudes still lead to ensembles of 
surrogates having a small variability of the PS. 
Algorithm B deal with this problem by allowing for a variability in the PS. 
The choice of the algorithm will depend upon the definition of the null 
hypothesis for every specific problem.\\ 
In view of our results, phase-controlled surrogate realizations are data sets 
which best reproduce the amplitude distribution in real space and the PS 
of the original data while explicitly fulfilling the constraint that only linear 
correlations are contained.

\bibliographystyle{ws-procs9x6}
\bibliography{surro_rand_phases}

\end{document}